\documentclass[aps,prb,showpacs]{revtex4}

\usepackage{mathptm}    
\usepackage{dcolumn}    
\usepackage{bm}         
\usepackage{graphicx}
\usepackage{times}

\begin{document}

\title{Two component spin-fermion model for high-$T_c$ cuprates:
Applications to neutron scattering and ARPES experiments}

\author{Yunkyu Bang}

\affiliation{Department of Physics, Chonnam National University,
Kwangju 500-757, Korea}

\begin{abstract}
Motivated by neutron scattering experiments in the high-$T_c$
cuprates, we propose the two-component spin-fermion model as a
minimal phenomenological model which has both local spins and
itinerant fermions as independent degrees of freedom. Our
calculations of the dynamic spin correlation function provide a
successful description of the puzzling neutron experiment data and
show that: (1) the upward dispersion branch of magnetic
excitations is mostly due to the local spin excitations; (2) the
downward dispersion branch is from collective particle-hole
excitations of fermions; and (3) the resonance mode is a mixture
of both degrees of freedom. Using the same model with the same set
of parameters we calculated the renormalized quasiparticle
dispersion and successfully reproduced one of the key features of
the angle resolved photoemission spectroscopy (ARPES) experiments,
i.e., the high energy kink structure in the fermion quasiparticle
dispersion, hence further support the two component spin-fermion
phenomenology.
\end{abstract}


\maketitle

\section{Introduction}
The study of spin dynamics has been a key research interest since
the discovery of the high-$T_c$ superconductor because it is
expected that the spin correlation holds crucial information for
the mechanism of the high-$T_c$ superconductivity (HTS). For long
the two main observations in the neutron scattering experiments of
high-$T_c$ cuprates (HTC) are (1) the incommensurate (IC) peaks at
low energy or at quasielastic excitations \cite{Cheong,Dai} and
(2) the so-called resonance peak at commensurate wave vector at
relatively high energy (30 $\sim$ 50 meV)
\cite{resonance,resonance2,resonance3}. In early experiments the
IC peaks were observed only in the deeply underdoped lanthanum
cuprate and the resonance mode was the hallmark of the fully doped
two layer yttrium cuprate. However, later experiments reveal that
both features appear in both groups of cuprate compounds, although
sensitively depending on doping level. More recently with
inelastic neutron scattering (INS)
experiemnts\cite{Arai99,Hayden04,Tranquada04,Pailhes04,Hinkov04,Keimer-condmat06}
with high precision, an unifying form of the magnetic excitations
in the cuprate superconductors has emerged as "hourglass" shape of
excitations around the wave vector (1/2,1/2) (hereafter in units
of $2 \pi / a$), in which the low energy IC excitations form the
downward dispersion branch and the high energy IC excitations form
the upward dispersion branch, and the two branches of excitations
merge at the commensurate momentum {\bf Q}=(1/2,1/2) and at the
resonance frequency $\Omega_{res}$.

It is a pressing question to understand the origin  of this
"hourglass" shape excitations. Theoretical proposals up to now can
be classified into two groups: (1) theories based on the spin
dynamics in the presence of stripes
\cite{stripes,stripes2,stripes3,stripes4}, and (2) Fermi liquid
type theories of itinerant fermions
\cite{FL-theories,FL-theories2,FL-theories3,FL-theories4,Sega}.
The key idea of the first group of theories is that the stripes formed by
doping in the two dimensional Cu-O plane splits the commensurate spin wave
excitations into two IC branches at the wave vectors (1/2 $\pm \delta$, 1/2)
or at their symmetry rotated positions by $ x \longleftrightarrow y$
depending on the directions of the stripes. The dispersions from each branch
of the two IC modulation cross at the commensurate wave vector (1/2,1/2) at
a higher energy, which is then identified as the resonance mode. This
picture provides a qualitative explanation to the hourglass dispersion and
the resonance mode. However, this type of theories has difficulty to be
extended to the higher doping regime where the presence and the nature of
the stripes is questionable.
The second group of proposals are itinerant fermion theories with
interaction \cite{FL-theories,FL-theories2,FL-theories3,
FL-theories4,Sega}. In this type of theories, the resonance mode
and the downward dispersion can be obtained, but the upward
dispersion branch is not yet satisfactorily reproduced.

In this paper, we propose a two component spin-fermion model
\cite{Bang,Bang2} as a minimal phenomenological model to provide a
natural and unifying explanation of the above mentioned neutron
experiments of HTC. In this phenomenological model, the minimal
set of low energy degrees of freedom are the spin wave excitations
of local spins and the continuum particle-hole excitations of
fermions.
A similar phenomenological theory is also known as one component
spin-fermion model and has been intensively studied by Pines and
coworkers \cite{Pines-review}. The main difference of the two
component model from the one component one is the introduction of
the spin wave excitations directly from the local spins in
addition to the usual collective spin density excitations from
fermions. In this paper, we show that the presence of the local
spin fluctuations is essentially proven by the INS and angle
resolved photoemission spectroscopy (ARPES) experiments of HTC,
therefore supporting the two component spin-fermion model as a
minimal phenomenological model of HTS.

\section{Formalism}
In a mixed momentum and real-space representation the two
component spin-fermion phenomenology Hamiltonian is written as

\begin{equation}
\label{eq.1}
 H = \sum_{{\bf k}, \alpha} c^{\dag}_\alpha({\bf
 k})\varepsilon({\bf k})c_\alpha({\bf
 k}) + \sum_{{\bf r},\alpha, \beta} g {\bf \vec{S}}({\bf r}) \cdot
c^{\dag}_\alpha({\bf r}){\bf \vec{\sigma}}_{\alpha \beta}c_\beta({\bf
 r}) + H_S ({\bf \vec{S}}({\bf r})),
\end{equation}

\noindent where the first term is the fermionic kinetic energy and
the second term describes the coupling between local spins  ${\bf
\vec{S}}({\bf r})$ and the spin density of the conduction
electrons ${\bf \vec{s}}({\bf r})= c^{\dag}_\alpha({\bf r}){\bf
\vec{\sigma}}_{\alpha \beta}c_\beta({\bf r})$. The last term $H_S
({\bf \vec{S}}({\bf r}))$ represents an effective low-energy
Hamiltonian for the local spins ${\bf \vec{S}}({\bf r})$. Instead
of specifying $H_S$ and solving it, we assumed a phenomenological
Ansatz of the bare (before coupling to the fermions) local spin
dynamics $<{\bf \vec{S}}_q {\bf \vec{S}}_{-q}>_{\Omega} =
\chi_{0,S}({\bf q}, \Omega)$ with a short range AFM correlation,
which has the general form as follows\cite{AFM}.
\begin{equation}
\label{eq.2} \chi_{0,S} ^{-1} ({\bf q}, \Omega) = \chi_{0,S}
^{-1}({\bf Q}, 0) \cdot [1 +  \xi^{2} |{\bf q} -{\bf Q}|^2 -
\Omega^2 / \Delta_{SG} ^2 ],
\end{equation}
\noindent where ${\bf Q}$ the 2D AFM ordering vector, and the spin
gap energy $\Delta_{SG}$ and the magnetic correlation length $\xi$
combine to give the spin wave velocity $v_s = \Delta_{SG} \cdot
\xi $ which can be determined by direct measurement\cite{Coldea}.

The key difference of the Hamiltonian Eq.(1) from the one
component Hamiltonian \cite{Pines-review} is the definition and
meaning of the spin fields ${\bf \vec{S}}({\bf r})$. We assumed
that ${\bf \vec{S}}({\bf r})$ is the local spin degrees of freedom
(d.o.f.) besides and independent from fermions
$c^{\dag}_\alpha({\bf r})$ and related itinerant spin density
${\bf \vec{s}}({\bf r})= c^{\dag}_\alpha({\bf r}){\bf
\vec{\sigma}}_{\alpha \beta}c_\beta({\bf r})$. On the other hand,
in one component spin-fermion model \cite{Pines-review}, the
collective spin fields is defined as the itinerant spin density
operator made of fermions, hence fundamentally linked to the
fermions, and there is no concept of the local spins. Another
important distinct feature of our two component phenomenology is
the local spin dynamics defined in Eq.~(\ref{eq.2}). This form of
the local spin correlation function with a short range AFM order
should be valid not only near ${\bf Q}$ but also for the entire BZ
of ${\bf q}$ -- if the precise form of the spin-wave dispersion is
ignored -- because at high energies the local spin dynamics
becomes less sensitive to the long range order or short range
order. However, the similar form of the itinerant spin correlation
function assumed in the one component model \cite{Pines-review} is
valid only in a narrow region of ${\bf q}$ around ${\bf Q}$ by
definition.

Microscopic justification of the above two component model,
starting, for example, from Hubbard or t-J model, is the heart of
problem of HTC for the last twenty years or so. We can only sketch
here the underlying idea for our phenomenology. Starting from a
Hubbard model, for example, dynamic mean field theory (DMFT)
\cite{DMFT} demonstrated that the key consequence of the strong
correlation of large $U$ Coulomb interaction is to split the
electron spectral density into two parts: one near the Fermi level
- the itinerant one, and the other at the lower and upper Hubbard
bands far below and above the Fermi level - hence the localized
one. Therefore this splitting of one bare electron spectral
density into the itinerant and localized parts is not a new
observation but has already had a solid theoretical justification.

A new step in our phenomenology is to propose that the localized
spectral density far away from the Fermi level is not dormant for
the low energy physics. In the framework of the DMFT, once the
coherent band is formed at the Fermi level in addition to the
upper and lower Hubbard bands through the strong correlation
effect, the low energy physics is solely described by the coherent
band near Fermi level and the Hubbard bands appear only as high
energy charge fluctuations such as the incoherent absorption bands
at high frequencies of order $O(U)$, for example, in optical
conductivity. This picture is correct with respect to the charge
degree of freedom because the DMFT is designed to capture the
strong correlation of charge dynamics by being a single site
impurity model. However, it is physically rather obvious that the
localized Hubbard bands can still contribute to low energy physics
through spin fluctuations. In order to capture this low energy
spin degree of freedom, it is, however, necessary to study the
lattice model -- not a small cluster but thermodynamically large
lattice. Then we have to give up all the merits of the DMFT. At
the moment, there is no satisfactory microscopic theory for the
lattice model which faithfully treats the strong correlation of
large $U$.

\begin{figure}
\hspace{0cm}
\includegraphics[width=130mm]{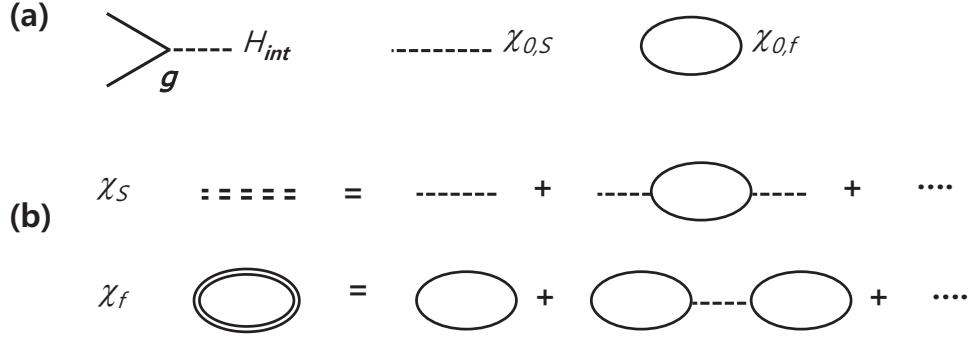}
\vspace{-3.5cm} \caption{(a) Interaction vertex of $H_{int}$, and
bare spin correlation functions $\chi_{0,S}$ and $\chi_{0,f}$. (b)
The graphic illustration of summations of the infinite series for
the dressed spin correlation functions $\chi_{S}$ and $\chi_{f}$
as defined in Eqs.(3) and (4)} \label{fig.1}
\end{figure}

To this end, we note that there exist two spin correlation
functions in our two component model: $\chi_S$ the one from the
local spins ${\bf \vec{S}}({\bf r})$ and $\chi_f$ the one from the
itinerant spin density ${\bf \vec{s}}({\bf r})=
c^{\dag}_\alpha({\bf r}){\bf \vec{\sigma}}_{\alpha
\beta}c_\beta({\bf r})$. Counting the coupling term to one loop
order (equivalent to the RPA), the dressed spin correlation
functions of the Hamiltonian (1) are written as follows.

\begin{eqnarray}
\label{eq.3} \chi^{-1} _{S} ({\bf q},\Omega) &=& \chi^{-1} _{0,S} ({\bf
q},\Omega) - g^2 \cdot \chi_{0,f} ({\bf q},\Omega) \\
\label{eq.4} \chi^{-1} _{f} ({\bf q},\Omega) &=& \chi^{-1} _{0,f} ({\bf
q},\Omega) - g^2 \cdot \chi_{0,S} ({\bf q},\Omega)
\end{eqnarray}

\noindent where $\chi_{0,S}$ is introduced in Eq.~(\ref{eq.2}) and
$\chi_{0,f}$ is the noninteracting spin susceptibility of the
conduction band of the fermions. The diagrammatic illustration of
the derivation of Eqs.(3) and (4) is shown in Fig.1. The
noninteracting spin susceptibility $\chi_{0,f}$ is written as

\begin{eqnarray}
\label{eq.5} \chi_{0, f} ({\bf q},\Omega) &=& \sum_k \frac{1}{2} \Big[1+
\frac{\epsilon({\bf k+q}) \epsilon({\bf k}) + \Delta({\bf k+q}) \Delta({\bf
k})}
{E({\bf k+q})E({\bf k})} \Big] \nonumber \\
&\times& \frac{f(E({\bf k+q}))-f(E({\bf k}))}
{\omega -[E({\bf k+q}) - E({\bf k})] +i \Gamma} \nonumber \\
&+& \sum_k \frac{1}{4} \Big[1 - \frac{\epsilon({\bf k+q}) \epsilon({\bf k})
+ \Delta({\bf k+q}) \Delta({\bf k})}
{E({\bf k+q})E({\bf k})} \Big] \nonumber \\
&\times& \frac{1 - f(E({\bf k+q}))-f(E({\bf k}))}
{\omega -[E({\bf k+q}) + E({\bf k})] +i \Gamma} \nonumber \\
&+& \sum_k \frac{1}{4} \Big[1 - \frac{\epsilon({\bf k+q}) \epsilon({\bf k})
+ \Delta({\bf k+q}) \Delta({\bf k})}
{E({\bf k+q})E({\bf k})} \Big] \nonumber \\
&\times& \frac{f(E({\bf k+q}))+f(E({\bf k}))-1} {\omega +[E({\bf k+q}) +
E({\bf k})] +i \Gamma},
\end{eqnarray}

\noindent where $E({\bf k})=\sqrt{\epsilon^2({\bf k})+\Delta^2({\bf k})}$
and the itinerant fermion dispersion $\epsilon({\bf k})$ is given by a tight
binding model

\begin{equation}
\label{eq.6} \epsilon ({\bf k}) = -2 t (\cos(k_x) + \cos(k_y)) - 2 t^{'}
\cos(k_x) \cdot \cos(k_y)-\mu.
\end{equation}

For calculations in this paper, we chose $t^{'}=-0.4 t$, and $\mu=
-0.81t$. The overall energy scale $t$ and the choice of parameters
$t^{'}$, $\mu$ will be discussed later with the numerical results.
For the superconducting state (SS), we assume a canonical d-wave
gap function $\Delta ({\bf k}) = \Delta_0 [ \cos(k_x) -\cos(k_y)]$
and for the normal state (NS), we set $\Delta_0 =0$ in Eq.(5).

Having two degrees of freedom in the model, two spin susceptibilities
$\chi_{S}$ and $\chi_{f}$ should be calculated on equal footing. Previous
studies of the local spin correlation embedded in the fermion bath
\cite{AFM} considered only the imaginary part of $\chi_{0,f}$ (so called
Landau damping) in Eq.~(\ref{eq.3}) to damp the spin wave excitations of
Eq.~(\ref{eq.2}) and the real part of $\chi_{0,f}$ is assumed either already
included in the definition of the bare local spin dynamics described in
Eq.~(\ref{eq.2}) or having negligible effects. In fact, when the coupling
$g$ is weak, this approach is reasonable. But in the strong coupling limit
when the dimensionless coupling constant $\lambda\equiv g^2 \cdot
\chi_{0,f}({\bf Q},0) \cdot \chi_{0,S}({\bf Q},0) \sim O(1)$, it is crucial
to include both the real and imaginary parts as in the above equations
(\ref{eq.3}) and (\ref{eq.4}). As we can see in the next section, in the
strong coupling limit both dressed spin susceptibilities $\chi_{S} ({\bf
q},\Omega)$ and $\chi_{f} ({\bf q},\Omega)$ become a mixture of the local
spins and the itinerant fermions and they assimilate to each other with
increasing the coupling strength $\lambda$.

In passing, the Eqs. (3) and (4) are loop expansions (one loop
order) but not a coupling constant expansion. Therefore, the
strong coupling limit of $\lambda \sim O(1)$ is not a problem, but
the higher loop diagrams -- for example, vertex corrections in a
standard many body terminology -- need to be worried. This
question is the beyond the scope of the current work.

\section{Neutron Scattering}

In order to study the INS experiments, we calculated the fully
dressed dynamic spin susceptibilities $ \chi_{S} ({\bf q},\Omega)$
and $\chi_{f} ({\bf q},\Omega)$ of Eqs. (3) and (4). As mentioned
in the previous section, in the strong coupling limit of $\lambda
\sim O(1)$, the behaviors of $ \chi_{S} ({\bf q},\Omega)$ and
$\chi_{f} ({\bf q},\Omega)$ become qualitatively similar each
other. Therefore, we  conveniently discuss the numerical results
of $\chi_{f} ({\bf q},\Omega)$ in this paper. But for
completeness, we also show the numerical results of $\chi_{S}
({\bf q},\Omega)$ as well as the total spin susceptibility
$\chi_{tot} = \chi_{f} + \chi_{S}$, too.

\begin{figure}
\hspace{0cm}
\includegraphics[width=130mm]{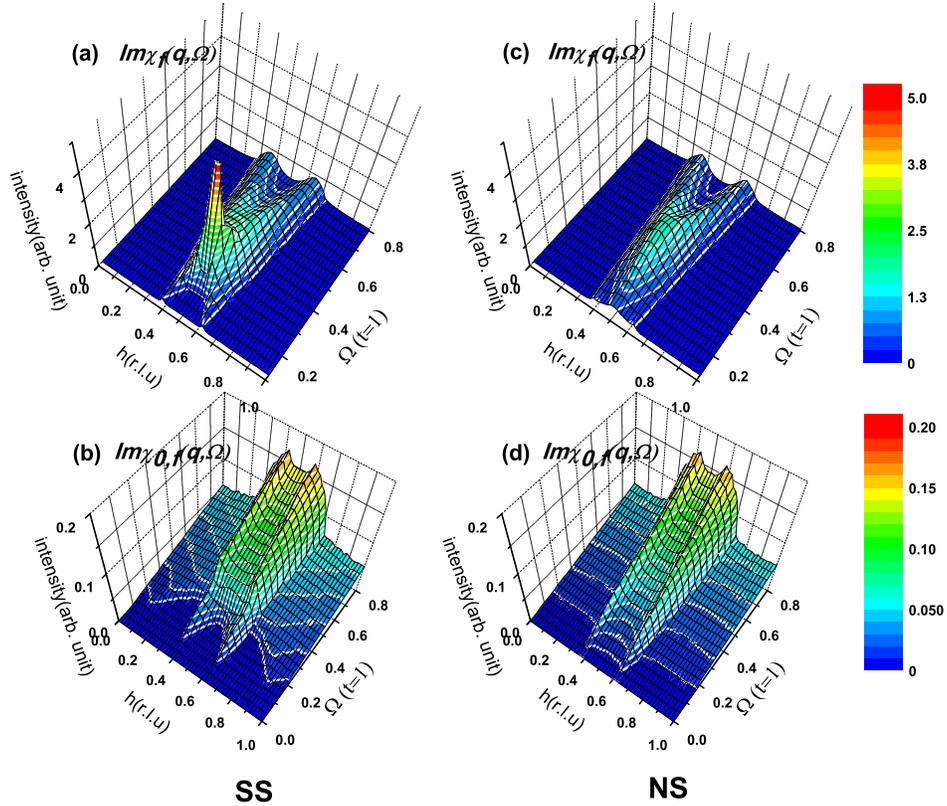}
\caption{(Color online) (a-b, in the left column) The dressed
itinerant spin susceptibility $Im \chi_{f}({\bf q}=(h,1/2),
\Omega)$ and the bare spin susceptibility $Im \chi_{0,f}({\bf q},
\Omega)$, respectively, in the superconducting state. Parameters
are $\Delta_{SG}=1.1 t$, $\Delta_0 = 0.2 t$, and $\lambda=0.8$;
(c-d, in the right column) $Im \chi_{f}({\bf q}, \Omega)$ and $Im
\chi_{0,f}({\bf q}, \Omega)$, respectively, in the normal state
($\Delta_0 =0$).} \label{fig.2}
\end{figure}

\subsection{$\chi_{f} ({\bf q},\Omega)$ along (0,1/2) $\rightarrow$ (1,1/2)}

Figure~\ref{fig.2}(a) shows $Im \chi_{f} ({\bf q},\Omega)$ scanned
along ${\bf q}= (h,1/2)$ in the SS. The superconducting gap
$\Delta_0 =0.2 t$, the bare spin gap $\Delta_{SG}=1.1 t$ (it is
not the physical spin gap), and the dimensionless coupling
constant $\lambda=0.8$ ($g^2=0.95$ eV$^2$) were chosen.
The main effect of the coupling is to renormalize down the bare
spin gap energy $\Delta_{SG}$ below the particle-hole excitation
gap of $\chi_{0,f} ({\bf q},\Omega)$ ($\sim 2 \Delta_0$), which
then forms a sharp resonance peak at ${\bf Q}=(1/2,1/2)$.
Centering from this resonance mode, both the downward dispersion
branch and the upward dispersion branch span out. The origin of
the upward dispersion is apparently from the local spin wave mode
(see Eq.~(\ref{eq.2})) and the origin of the downward dispersion
is the itinerant spin excitations of $\chi_{0,f}$. The latter fact
can be identified in Fig.~\ref{fig.2}(b) which shows the
non-interacting fermion spin susceptibility $Im \chi_{0,f} ({\bf
q},\Omega)$ scanned along ${\bf q}= (h,1/2)$ in the SS. The shape
and strength of the downward whisker like excitations in $Im
\chi_{0,f}$ is sensitive to the Fermi surface (FS) curvature
(controlled by $\mu$ and $t^{'}$), and the size of the d-wave gap
$\Delta ({\bf k}) = \Delta_0 [ \cos(k_x) -\cos(k_y)]$.

With the coupling strength $\lambda=0.8$, the dressed fermion spin
susceptibility $\chi_{f} ({\bf q},\Omega)$ obtains features of
both the local spin susceptibility $\chi_{0,S}$ and the itinerant
spin susceptibility $\chi_{0,f}$. In particular, the high energy
parallel branches in $\chi_{0,f}$ (see Fig.~\ref{fig.2}(b) and
(d)) are overwhelmed by the spin wave like excitations of
$\chi_{0,S}$ in the dressed susceptibility $\chi_{f}$ as seen in
Fig.~\ref{fig.2}(a) and (c). With a smaller coupling strength
($\lambda < 0.5$) the two spin susceptibilities $\chi_{S} ({\bf
q},\Omega)$ and $\chi_{f} ({\bf q},\Omega)$ retain more of their
bare characteristics of the spin wave excitations and the
itinerant fermion susceptibility, respectively, and the resonance
peak at ${\bf Q}=(1/2,1/2)$ is not formed.

\begin{figure}
\hspace{0cm}
\includegraphics[width=130mm]{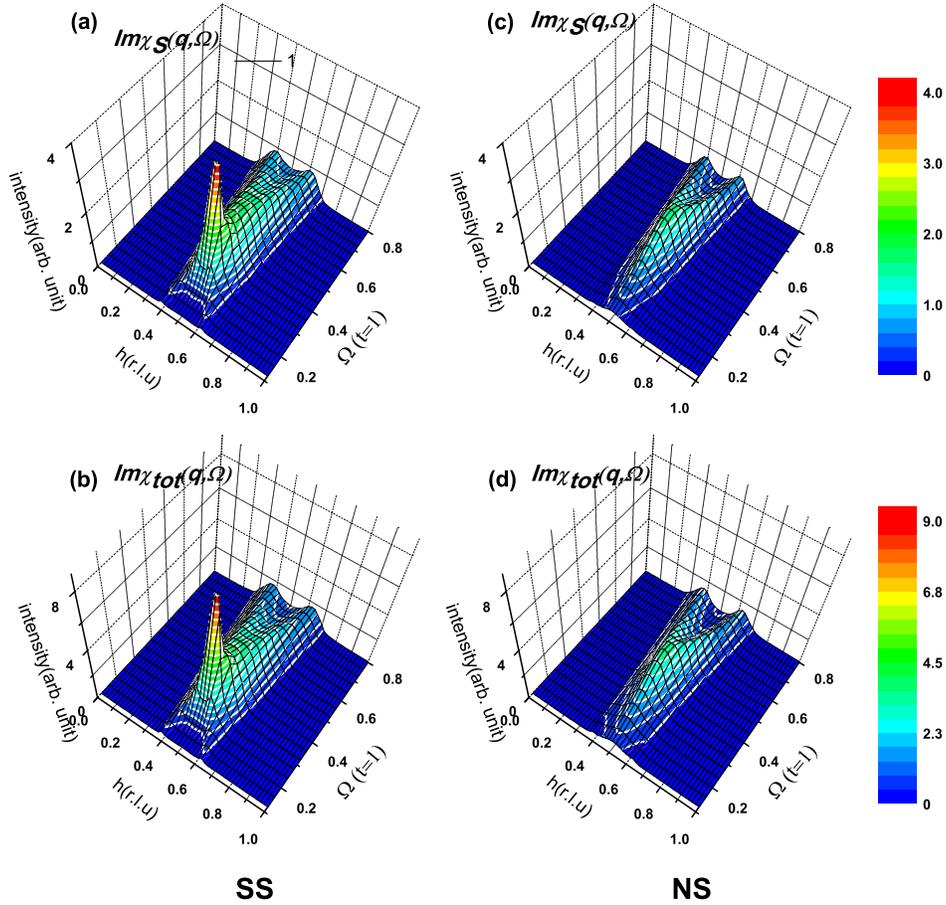}
\caption{(Color online) (a-b, in the left column) The dressed
local spin susceptibilities $Im \chi_{S}({\bf q}=(h,1/2), \Omega)$
and the total spin susceptibilities $Im \chi_{tot}({\bf
q}=(h,1/2), \Omega)= Im \chi_{S} +Im \chi_{f}$, respectively, in
the superconducting state. (c-d, in the right column) $Im
\chi_{S}({\bf q}, \Omega)$ and $Im \chi_{tot}({\bf q}, \Omega)$,
respectively, in the normal state. Parameters are the same as in
Fig.2.} \label{fig.3}
\end{figure}

Figure~\ref{fig.2}(c,d) are the same plots as in
Fig.~\ref{fig.2}(a,b) but in the NS. First, the resonance peak
becomes severely overdamped having only a hump like structure in
$Im \chi_{f} ({\bf q},\Omega)$. Second, the low energy downward
whisker like dispersion disappears because the free fermion
susceptibility $\chi_{0,f} ({\bf q},\Omega)$ in the NS (see
Fig.~\ref{fig.2}(d)) has no such structure. Lastly, the high
energy upward dispersion remains almost similar to the case of the
SS.
The results of Fig.~\ref{fig.2}(a,c) successfully reproduce the
main features of recent neutron scattering experiments in HTC
\cite{Arai99,Hayden04,Tranquada04,Pailhes04, Hinkov04,
Keimer-condmat06}, ie., the resonance mode in SS, the hourglass
shape of the upward and downward dispersions, and their drastic
change between superconducting and normal states. In particular,
these results strikingly resemble the INS data of YBCO$_{6.6}$
\cite{Keimer-condmat06}.
%
However, we need a reservation for applying our result to
L$_{1.875}$B$_{x=0.125}$CO$_{4}$ \cite{Tranquada04}, which has a
static stripe order and extremely low T$_c= 2.5$ K
\cite{Valla-ARPES}. In particular, the presence of the stripe
ordering is likely to change the spin dynamics significantly
\cite{stripes2,stripes3,stripes4} and introduce the a-b plane
anisotropy.

For comparison, we also show the numerical results of the local
spin susceptibility $\chi_S ({\bf q},\Omega)$ as well as the total
spin susceptibility $\chi_{tot} ({\bf q},\Omega) = \chi_f ({\bf
q},\Omega) + \chi_S ({\bf q},\Omega)$ in Fig.~\ref{fig.3}. As
mentioned above, the overall behavior of $\chi_S ({\bf q},\Omega)$
is indistinguishably similar to $\chi_f ({\bf q},\Omega)$ and so
is $\chi_{tot} ({\bf q},\Omega)$. This is the typical feature of
the strong coupling limit of $\lambda \sim O(1)$ and remember that
the value of $\lambda=0.8$ used in our calculations was not an
arbitrary choice but was determined by the $(\pi, \pi)$ resonance
condition. As expected, however, a fine difference exists so that,
in general, $\chi_S ({\bf q},\Omega)$, in comparison to $\chi_f
({\bf q},\Omega)$, has a slightly more feature of the local spin
dynamics at higher frequencies and a slightly less feature of the
itinerant fermion spin dynamics at low frequencies and vice versa.
For example, we can see a bit weaker downward dispersion branch in
$\chi_S ({\bf q},\Omega)$ in the SS (Fig.3(a)) than in $\chi_f
({\bf q},\Omega)$ in the SS (Fig.2(a)).

Also although we plotted the total spin susceptibility as
$\chi_{total} =\chi_f + \chi_S$, there is an ambiguity about
whether the contributions from the local spin fluctuations
$\chi_S$ and the itinerant spin fluctuations $\chi_f$ to the INS
measurement should be equal as we tentatively assumed here because
the form factors of the local and itinerant spins, in principle,
should be different. However, our ignorance of the relative
strength of the $\chi_f$ and $\chi_S$ fluctuations does not affect
our phenomenology because the physical coupling strength between
two spin fluctuations is determined by the effective dimensionless
coupling $\lambda (q) \equiv g^2 \cdot \chi_{0,f}({\bf q},0) \cdot
\chi_{0,S}({\bf q},0)$ and the value of $\lambda ({\bf Q})$ is
determined once and for all by the $(\pi, \pi)$ resonance
condition and all other physical quantities are calculated without
further ambiguity. Keeping this point in mind, the comparison of
our numerical calculations to the INS experiments should be
qualitatively the same whether we use the results of $\chi_f$ or
$\chi_S$, or $\chi_{tot}$.

\begin{figure}
\hspace{2cm}
\includegraphics[width=140mm]{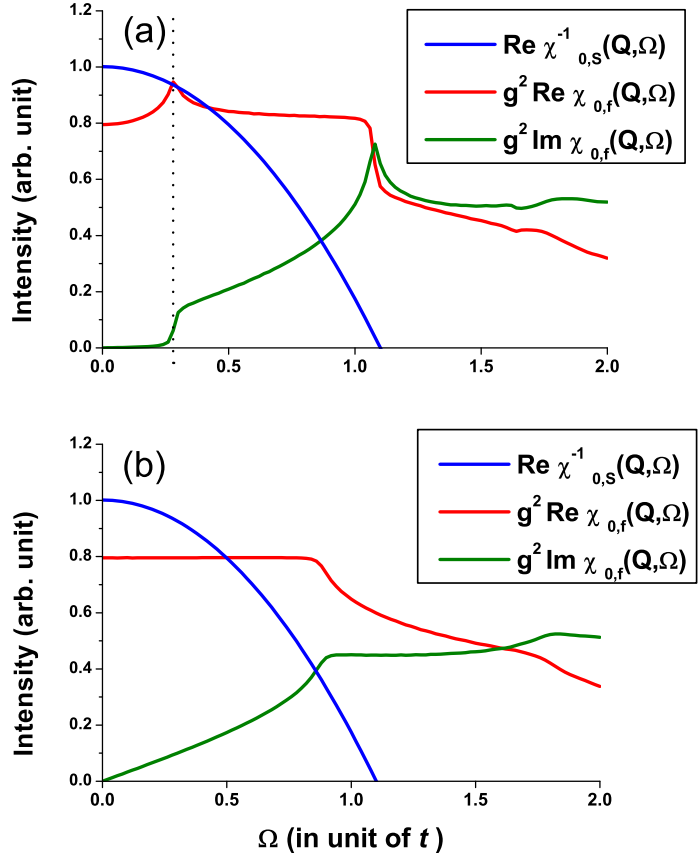} \caption{(Color online) (a) Plots
of bare susceptibilities $Re \chi^{-1}_{0,S}({\bf Q},\Omega)$,
$g^2 Re \chi_{0,f}({\bf Q},\Omega)$, and $g^2 Im \chi_{0,f}({\bf
Q},\Omega)$, respectively in superconducting state. Parameters are
$\Delta_{SG}=1.1 t$, $\Delta_0 = 0.2 t$, and $\lambda=0.8$. The
vertical dashed line is a guide to the eyes indicating the
position of pole at $\Omega=0.28 t$. (b) The same as (a) in normal
state (ie. $\Delta_0 =0$).} \label{fig.4}
\end{figure}

\subsection{Origin of the resonance mode}

The mechanism of forming the resonance mode in our model is
illustrated in Fig.~\ref{fig.4}.  When the inverse of the dressed
susceptibilities of Eq.~(\ref{eq.3}) and Eq.~(\ref{eq.4}) crosses
zero (which occurs simultaneously in both susceptibilities), the
dressed susceptibilities develop a resonance mode: a bound state
or an overdamped mode depending on the presence and strength of
the imaginary part at the position of pole.
In Fig.~\ref{fig.4} we plot separately $Re \chi^{-1} _{0,S} ({\bf
Q},\Omega)$, $Re \chi_{0,f} ({\bf Q},\Omega)$, and $Im \chi_{0,f}
({\bf Q},\Omega)$ to make this point clear.
Figure~\ref{fig.4}(a) is the case of a SS, where the pole of
$\chi_{f,S} ({\bf q =\bf Q},\Omega)$ occurs at $\Omega_{res} \sim
0.28 t$. At this frequency the damping from $Im \chi_{0,f}$ is
very weak below the p-h excitation gap, so that the pole becomes a
sharp resonance peak.
Figure~\ref{fig.4}(b) shows the case of the NS ($\Delta_0 =0$)
with the same parameters as in Fig.~\ref{fig.2}(c). The position
of pole occurs at a little higher frequency ($\Omega \sim 0.5 t$)
compared to the case of the SC phase (Fig.~\ref{fig.4}(a)). But
this pole is strongly damped by $Im \chi_{0,f}$ (green line) that
is linearly increasing with energy, and this linearly increasing
damping shifts down the actual position of the peak to
$\Omega_{res} \sim 0.35 t$ (the maximum height position in
Fig.2(c)). We note that this overdamped resonance peak at NS is
consistent with the data of Ref\cite{Keimer-condmat06} as shown in
Fig.~\ref{fig.2}(c).

The resonance mode found in our model has physically different
content than the resonance mode in the Fermi liquid type theories
\cite{FL-theories,FL-theories2,FL-theories3, FL-theories4}. The
line of $Re \chi^{-1} _{0,S} ({\bf Q},\Omega)$ in Fig.~\ref{fig.4}
is not a simple inverse of a static potential (for example,
$\frac{1}{U(q)}$ in a RPA calculation of Hubbard model as in
\cite{FL-theories2,FL-theories3, FL-theories4}) but it carries its
own dynamics and spectral density. Therefore, the resonance mode
formed by coupling of two dynamic susceptibilities $\chi_{0,f}$
and $\chi_{0,S}$ should carry the spectral densities from both the
local spin wave and the fermion particle-hole continuum. In our
coupled two component spin-fermion model, the upward excitation
branch and the resonance mode appear from a pole of
Eq.~(\ref{eq.3}) and Eq.~(\ref{eq.4}) for a given ${\bf q}$ but
the downward excitation branch is made of particle-hole
excitations of fermions in the d-wave SS and does not constitute a
pole in Eq.~(\ref{eq.3}) and Eq.~(\ref{eq.4}). This is in contrast
with the Fermi liquid type theories
\cite{FL-theories,FL-theories2,FL-theories3,FL-theories4,Sega}
where both the downward branch and the resonance mode are
constructed by the pole of a RPA type spin susceptibility.

\subsection{Constant energy scans}

\begin{figure}
\hspace{0cm}
\includegraphics[width=190mm]{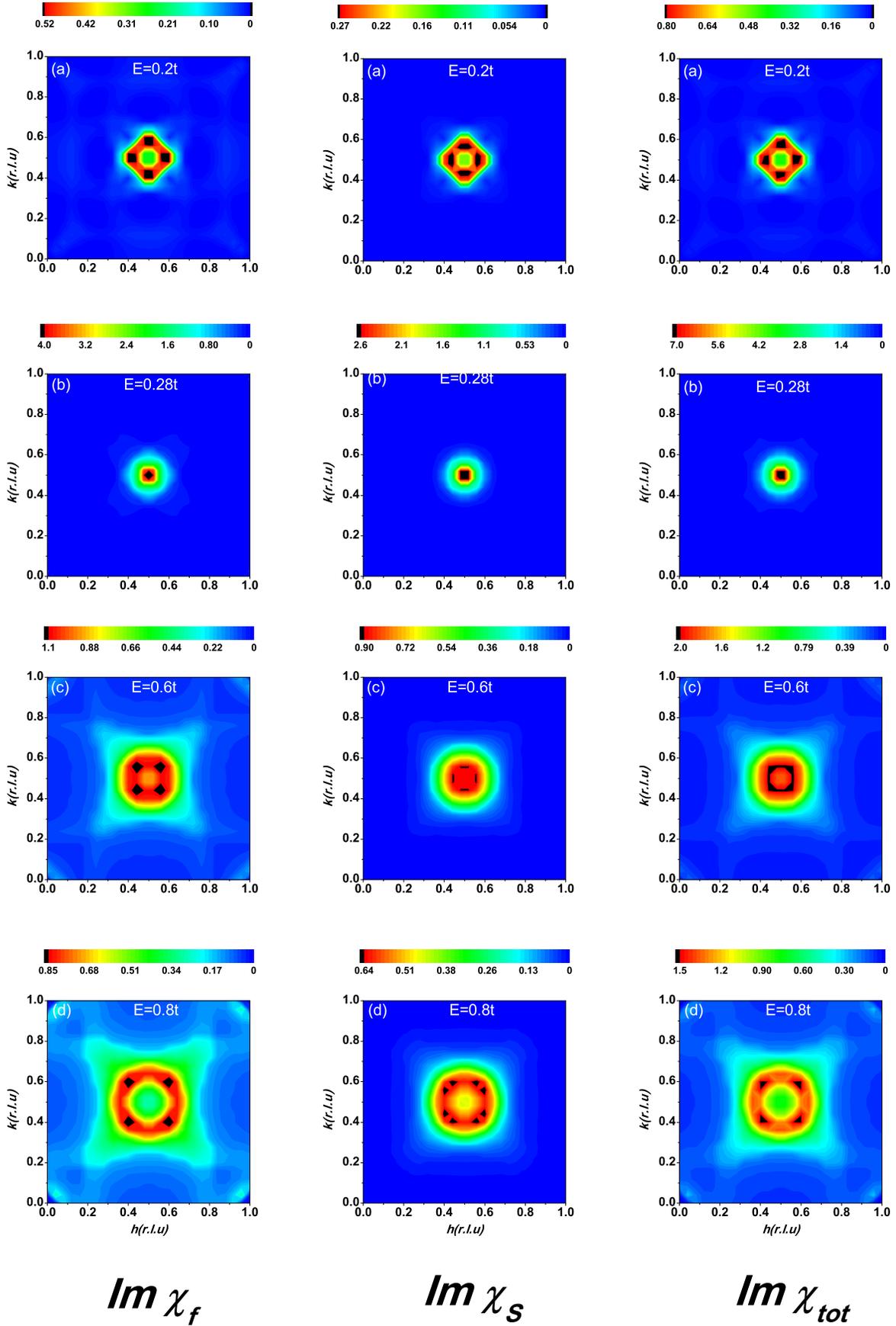} \vspace{-2.0cm} \caption{(Color
online) Constant energy scans of $Im \chi_{f}({\bf q},\Omega)$
(left column), $Im \chi_{S}({\bf q},\Omega)$ (center column), and
$Im \chi_{tot}({\bf q},\Omega)$ (left column) at (a) $\Omega=0.2
t$, (b) $\Omega=0.28 t$, (c) $\Omega=0.6 t$, and (d) $\Omega=0.8
t$, respectively. In all cases, parameters are $\Delta_{SG}=1.1
t$, $\Delta_0 = 0.2 t$, and $\lambda=0.8$.} \label{fig.5}
\end{figure}

In the left column of Fig.~\ref{fig.5}.(a-d), we show the constant
energy scans of $\chi_{f} ({\bf q},\Omega)$ in the SS for
$\Omega=0.2t, 0.28t, 0.6t$, and $0.8t$, respectively.
Constant energy scans of neutron scattering data of YBCO
\cite{Hayden04} and LBCO \cite{Tranquada04} show peculiar patterns
of IC peak positions in $(q_x, q_y)$ momentum space at different
energy cuts. In particular, the 45 $\deg$ rotation of the patterns
from a low energy scan (below the resonance energy $\Omega_{res}$)
to a high energy scan drew special attention and several
theoretical explanations have been proposed
\cite{stripes2,stripes3,stripes4,FL-theories2,FL-theories3,FL-theories4}.
Results of Fig.~\ref{fig.5} demonstrate that the two component
spin fermion model can consistently explain this phenomena, too.

\hspace{-0cm}

Figure~\ref{fig.5}(b) in the left column is the scan of $\chi_{f}
({\bf q},\Omega)$ at the resonance energy, $\Omega_{res}=0.28 t$
with the same parameters as in Fig.~\ref{fig.2}(a). It shows a
very intense peak at (1/2,1/2) indicating a very sharp resonance
not only in energy but also in momentum space. Fig.~\ref{fig.5}(c)
and Fig.~\ref{fig.5}(d) are the scans at higher energies than the
resonance energy and Fig.~\ref{fig.5}(a) is a scan of lower energy
cut. We colored the highest intensity positions with black color
to emphasize the clear patterns. The lower energy scan
(Fig.~\ref{fig.5}(a)) shows the IC peaks at $(1/2 \pm \delta,
1/2)$ and $(1/2, 1/2 \pm \delta)$ forming a diamond shape pattern.
The higher energy scans (Fig.~\ref{fig.5}(c) and
Fig.~\ref{fig.5}(d)) show that the IC peak positions at $(1/2 \pm
\delta, 1/2 \pm \delta)$ and $(1/2 \pm \delta, 1/2 \mp \delta)$
forming a square shape pattern which has the symmetry of the 45
$\deg$ rotated from the low energy pattern. The results of
Fig.~\ref{fig.5} excellently reproduce the observed patterns of
the constant energy scan data of neutron experiments reported in
YBCO \cite{Hayden04} and LBCO \cite{Tranquada04}. This is rather
surprising for LBCO since this compound is known to develop a
static stripe ordering and our model has no ingredient for the
stripes as we mentioned before.

In our model we can trace the origins of the IC peak patterns. The low
energy IC peaks and diamond shape pattern is basically a reflection of the
band structure and d-wave superconducting gap. The high energy IC peaks and
the square shape pattern has more complicated origin. At and above the
resonance energy the dressed spin susceptibility $\chi_{f}$ is the result of
a strong interplay between the local spin correlation and the itinerant spin
correlation. Therefore the high energy scan pattern is the result of a
subtle interplay/competition between $\chi_{0,S}({\bf q},{\Omega})$ and
$\chi_{0,f}({\bf q},{\Omega})$.
The presence of IC peaks at high energies itself is the
manifestation of the high energy spin wave dispersion spanning
from the AFM wave vector ${\bf Q}$; so the incommensurability
increases with energy. However, whether the pattern will be a
square or a diamond shape has no universal mechanism.
We tested various combinations of parameters $t^{'}$, $\mu$,
$\Delta_{SG}$, $\Delta_{0}$ and $\lambda$. The low energy diamond
shape pattern is robust within our model. As to the patterns of
higher energy scans, although the square shape is the dominant
one, it is not absolutely robust; with different parameters the
diamond pattern can appears, too. Therefore, we think that the 45
$\deg$ rotation of the IC peak patterns may not be an universal
feature of HTC; it can change with doping and for different
cuprate compounds. This non-universality is also seen in the scan
of $\chi_{S}({\bf q},{\Omega})$ at $\Omega_{res}=0.6 t$ in the
center column of Fig.~\ref{fig.5}(c). However, as repeatedly
emphasized, the patterns of the constant energy scans for
$\chi_{S}({\bf q},{\Omega})$, $\chi_{tot}({\bf q},{\Omega})$, and
$\chi_{f}({\bf q},{\Omega})$ are basically the same each other in
the strong coupling limit.
\subsection{Parameters of our phenomenology}

To make a comparison of our calculations with experiments, it is important
to fix the energy scale of the model. The tight binding band of
Eq.~(\ref{eq.6}) is widely studied to fit the ARPES data and the estimate of
$t$ varies from 150 meV to 400 meV depending on the doping and different
cuprate compounds \cite{band}. Our calculation results are in good agreement
with neutron experiments in terms of energy scale if we choose $t \sim$
150-180 meV. This value of $t$ corresponds to the low end of the estimates
from ARPES experiments. One possible reason for it is that the extraction of
$t$ value from ARPES is carried by fitting the whole Brillouin zone (BZ) of
the quasiparticle (q.p.) dispersions. As a result the high energy dispersion
sets the overall energy scale $t$. However, the low energy spin
susceptibility is determined by the low energy particle-hole excitations
near Fermi level and irrelevant with the high energy q.p. excitations.
With this reasoning it is quite possible that the effective $t$ value near
FS is in fact much reduced by a renormalization due to the strong
correlation effect.

The degree of incommensurability and the strength of the downward
whisker-like dispersion in the fermion susceptibility $\chi_{0,f}$
(see Fig.~\ref{fig.2}(b)) is controlled by the FS curvature --
which is tuned by $t^{'}$ and the chemical potential $\mu$ -- and
the SC gap size $\Delta_0$; this property is true even for the
Fermi liquid theories or one component spin-fermion theories
\cite{FL-theories,FL-theories2,FL-theories3,FL-theories4}. These
parameters $t^{'}$, $\mu$, and $\Delta_0$ should be independently
determined by other experiments such as ARPES\cite{band},
tunneling\cite{tunneling}, etc. Therefore, after fixing the
overall energy scale of the model by $t$, the genuinely free
fitting parameters of our phenomenological model are only two: the
coupling strength $g$ (or equivalently $\lambda$) and the bare
spin gap $\Delta_{SG}$. For all calculations in this paper we used
$t^{'}=-0.4 t$, $\mu= -0.81t$, $\Delta_{0}=0.2t$,
$\Delta_{SG}=1.1t$, and $\lambda$=0.8.

\begin{figure}
\includegraphics[width=150mm]{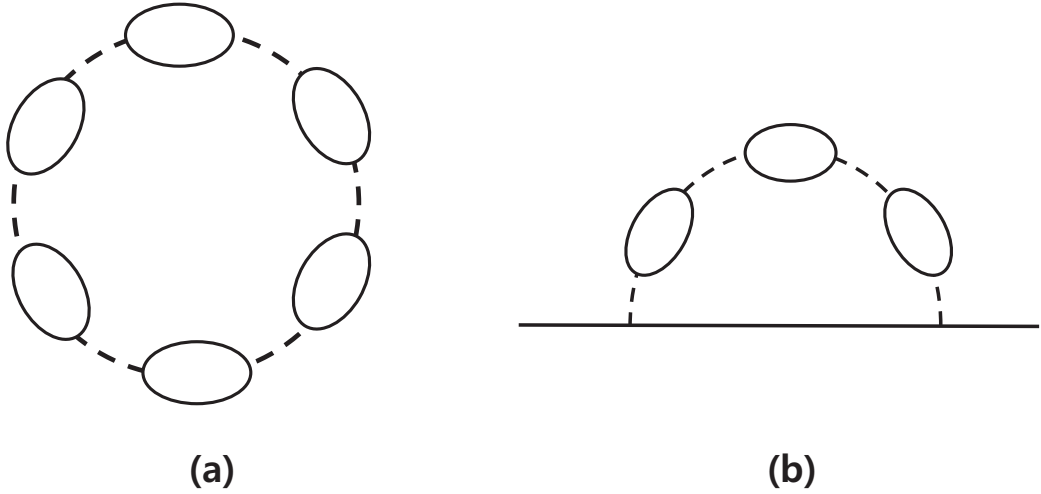}
\hspace{0.0cm} \vspace{-3.5cm} \caption{(a) A typical diagram of
the Free energy of the model Hamiltonian $H$ of Eq.(1) in one loop
approximation. (b) A typical diagram of the fermion selfenergy in
one loop approximation calculated in Eq.(7). The diagram (b) can
be obtained by cutting any fermion line of the diagram (a).}
\label{fig.6}
\end{figure}

\section{ARPES and high energy kink}

As a consistent check for our phenomenological model, we calculate
the renormalized band dispersion with the same parameter set that
we have used for the neutron scattering in previous section. In
this calculation, the most important parameter is the coupling
strength between fermions and spin fluctuations, for which there
is no direct experimental measurement nor a reliable theoretical
estimate from a microscopic Hamiltonian. In our phenomenology,
this value $\lambda\equiv g^2 \cdot \chi_{0,f}({\bf Q},0) \cdot
\chi_{0,S}({\bf Q},0) =0.8$ was determined by the condition to
produce $(\pi, \pi)$ resonance mode both in the superconducting
and normal states. So we can crosscheck the consistency of our
phenomenology by comparing the outcomes of the q.p.
renormalization from the spin-fermion interaction to the ARPES
experiments.

The selfenergy of the fermion q.p. is calculated in Born
approximation with the fully dressed local spin fluctuations
$\chi_S (q,\omega)$ as

\begin{equation}
\label{eq.7} \Sigma(\vec{k},\omega)= g^2 \sum_{q}  \int \frac{d
\omega'}{\pi} \frac{Im \chi_S (q,\omega')} {\omega +  \omega' -
\epsilon_{k+q} + i \Gamma} [n (\omega') + f(\epsilon_{k+q})]
\end{equation}
\noindent where $n (\omega)$ and $f (\omega)$ are the Boson and
Fermion distribution functions, respectively. Notice that this
calculation of the fermion selfenergy is the same one loop
approximation as the calculations of the spin susceptibility
renormalization in Eqs.(3) and (4), guaranteeing the consistency
of our phenomenology. This is graphically demonstrated in
Fig.~\ref{fig.6}. We then calculate the renormalized q.p. spectral
density as $A (\vec{k},\omega)= Im G_R (\vec{k},\omega)$ with $G_R
(\vec{k},\omega)=
\frac{1}{\omega-\epsilon_k-\Sigma(\vec{k},\omega)}$.

\begin{figure}
\includegraphics[width=150mm]{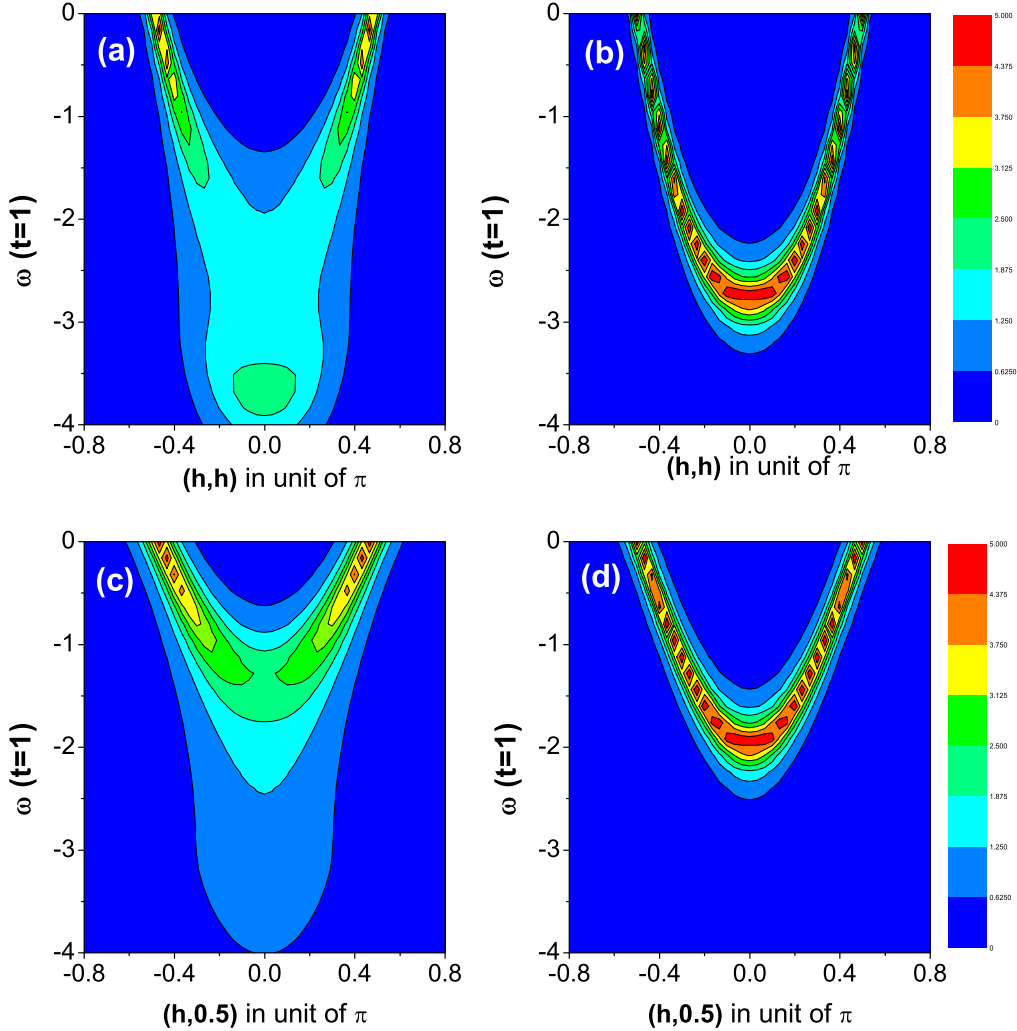}
\caption{(Color online) Spectral densities of quasiparticle
dispersions: (a, c) Renormalized band dispersions along (h,h) and
(h, 0.5) cuts, respectively. (b, d) Bare band dispersions along
(h,h) and (h, 0.5) cuts, respectively.} \label{fig.7}
\end{figure}

In Fig.~\ref{fig.7}, we show the contour plots showing the
intensity of the q.p. spectral densities with and without the
selfenergy correction. Fig.~\ref{fig.7}(a-b) show the dispersions
along the nodal direction $(0,0) \rightarrow (\pi, \pi)$ and
Fig.~\ref{fig.7}(c-d) show the dispersions along the near
antinodal direction $(0,0.5\pi) \rightarrow (\pi, 0.5\pi)$.
Two features are distinctively seen as results of the spin-fermion
interaction: (1) the overall q.p. dispersion is renormalized by a
factor of $\sim O(2)$. This is consistent with the input
$\lambda=0.8$ because the wave function renormalization factor is
$Z \approx (1 + \lambda)$; (2) much interesting point is that the
q.p.s with a certain energy below the Fermi level are so strongly
renormalized that the bare dispersion is not continuously
renormalized but is detached from the low energy dispersion and
forms a broad spectral puddle at further high energy region (see
Fig.~\ref{fig.7}(a) and (c)).

This second feature is commonly observed both for the nodal and
antinodal directions regardless of some differences of the fine
details such as the breaking points of dispersion in energy and
momentum space and the intensity of the broad spectral puddle at
high energies. This behavior is remarkably similar to the
so-called high energy kinks ($\sim 340 meV$) observed in the ARPES
experiments with BSCCO and LBCO by T. Valla {\it et al}
\cite{Valla-ARPES}
The breaking point of the dispersion occurs around $\omega \sim 1.5 - 2 t$
in our model calculations. In fact, if we assume $t \sim 150 -180 meV$ as
discussed before, the kink energy we calculated corresponds to $\sim 225 -
360 meV$, consistent with the experimental data. Our results even reproduce
the overall differences of the dispersion and kink behavior between the
nodal direction and the antinodal direction as observed in experiments.
\cite{Valla-ARPES}

We can trace the origin of this jump or breaking of the q.p.
dispersion and found that it is caused by the upper bound of the
local spin wave excitations $\chi_S ({\bf q},\Omega)$. The local
spin wave excitations, defined in Eqs.(2) and (3), disperses from
the lowest energy $\tilde{\Delta}_{SG}$ (renormalized one) at
${\bf q} = {\bf Q} =(\pi, \pi)$ to the highest energy at the
magnetic zone corners ${\bf q} = (2 \pi, 2 \pi)$ and its
equivalent points. Even after dressed by fermions as in Eq.(3)
this damped spin wave excitations has a similar upper bound. With
the parameters of our model, this high energy upper bound of the
local spin fluctuations is limited at around $2t$. Physical
meaning of it is that the local spin excitations exist only up to
$\sim 2t$ and the fermion q.p. cannot be scattered beyond this
energy scale. Consequently the real part of selfenergy calculated
with Eq.(\ref{eq.7}) develops a rapid variation at around $2t$ in
frequencies and q.p. pole is not formed beyond this energy scale
in the dressed fermion Green's function $G_R (\vec{k},\omega)=
\frac{1}{\omega-\epsilon_k-\Sigma(\vec{k},\omega)}$. Hence we can
understand the origin of the high energy kink from the high energy
upper bound of the local spin excitations \cite{hi-kink}.
Consistency between our model calculations and experimental
observation of the high energy kink strongly supports the
existence and strength of local spin excitations which have a
upper energy scale around $350 meV$. In contrast, the itinerant
spin fluctuations has a long tail of the particle-hole continuum
excitations up to the band width ($\sim 8 t$) which would be a
couple of $eV$ at least. Therefore, the one component spin-fermion
model with only the itinerant fermions would have a difficulty to
explain the high energy kink phenomena.

The reproduction of the high energy kink feature in the
renormalized q.p. dispersion also implies that not only the low
energy spin excitations near $(\pi, \pi)$ plays an important role
but also the high energy spin excitations can play an important
role. However, it doesn't necessarily mean that the coupling of
the high energy spin excitations to fermions has a comparable
strength as the coupling of the low energy spin excitations to
fermions. In fact, the effective coupling between fermions and the
local spins $\lambda({\bf q})$ becomes much weaker with ${\bf q}$
away from ${\bf Q}=(\pi,\pi)$. However, the phase space of the
local spin excitations rapidly increases with increasing ${\bf q}$
from ${\bf Q}=(\pi,\pi)$, which compensates for the weakness of
coupling.

Within our resolution, we didn't find noticeable features in the
low energy q.p. dispersion, i.e., the low energy kink, which might
be related with the $(\pi, \pi)$ resonance mode. Possibly it is
because the spectral weight of the resonance mode in NS (see
Fig.~\ref{fig.3}(c)) is not sufficiently dominant over the total
spectral density of the spin fluctuations  spread over the whole
momentum space, and/or more possibly because the low energy kink
is in fact not an abrupt kink but rather a gentle variation of the
dispersion slope as seen in the recent experiment \cite{Dahm}.

\section{Conclusion}
In summary, we proposed a phenomenological two-component spin
fermion model motivated by the neutron scattering experiments in
HTC. With the two spin degrees of freedom of the local spins and
the itinerant spins, our calculations of the dynamic spin
susceptibilities coherently reproduced the essential features of
the neutron experiments in HTC: the hourglass dispersions,
resonance mode, their changes in normal and superconducting
states, and the IC peak patterns of constant energy scans.
Although our approach is a phenomenology, considering that there
are genuinely only two free fitting parameters, i.e., the coupling
constant $\lambda$ and the bare spin gap $\Delta_{SG}$, the
successful reproduction of the several key features of neutron
experiments with one set of parameters is quite encouraging.
Then with the same model parameters, we calculated the
renormalized fermion q.p. dispersion and reproduced both the nodal
and antinodal high energy kinks with the correct energy scale in
agreement with ARPES experiments \cite{Valla-ARPES}. It further
strengthened the justification for our phenomenology.

Finally, the main message of this work with the two component
spin-fermion phenomenology is to demonstrate that there are
compelling experimental evidences\cite{Tacon} for the presence and
its important role of the local spin degrees of freedom in
addition to the fermionic quasiparticles in the cuprates.
Interestingly, there are also accumulating experimental evidences
for the coexistence of the itinerant electrons and local moment of
spins in the recently found iron-based superconducting compounds
\cite{Dual ARPES, Dual Neutron}. The pressing question is now what
the microscopic theory is for the phenomenological two-component
spin fermion model; in other words how the local spin degrees of
freedom survives doping from the parent insulating cuprate
compounds, or more generally how the local moments and the
itinerant fermions coexist in the strongly correlated metallic
systems such as cuprate and pnictide compounds.

The author (Y.B.) was supported by the Grant No. NRF-2010-0009523
and NRF-2011-0017079 funded by the National Research Foundation of
Korea.

\end{document}